\documentclass[preprint,aps,12pt,preprintnumbers,eqsecnum,nofootinbib]{revtex4}
\usepackage{graphicx}
\usepackage{subfigure}

\usepackage{color}
\usepackage{amssymb,amsmath}

\begin{document}
%
%
\title{\vspace*{0.5in} Technicolor with a 125 GeV Higgs Boson
\vskip 0.1in}
\author{Christopher D. Carone}\email[]{cdcaro@wm.edu}
\affiliation{High Energy Theory Group, Department of Physics,
College of William and Mary, Williamsburg, VA 23187-8795}
\date{June 2012}
\begin{abstract}
Bosonic technicolor models accommodate fermion masses via a Higgs doublet that acquires 
a vacuum expectation value when technifermions condense.  We point out that these models are 
severely constrained by vacuum stability if the Higgs boson mass is near 125 GeV, the value 
suggested by LHC data.  The Higgs quartic coupling in bosonic technicolor is typically smaller at 
the weak scale than in the Standard Model, while the top quark Yukawa coupling is larger.  We find 
that the running quartic coupling remains positive below a reasonably defined cut off only in
a narrow region of the model's parameter space.  This region is only slightly enlarged 
if one allows a metastable vacuum with a lifetime longer than the age of the universe. 
\end{abstract}
\maketitle

\section{Introduction} \label{sec:intro}

The simplest technicolor models achieve electroweak symmetry breaking via a 
condensate of fermions that are charged under a new, strong gauge group~\cite{technicolor}. If 
the LHC confirms the existence of a Higgs boson near $125$~GeV~\cite{h125} with couplings 
similar to those expected in the Standard Model, then the simplest technicolor models 
will be conclusively excluded, independent of the already powerful, albeit indirect, constraints 
from precision electroweak  measurements~\cite{PT}.

This observation, however, does not  preclude the possibility that new strong dynamics 
might contribute {\em in part} to the breaking of electroweak symmetry.   Bosonic technicolor 
models provide an example of this scenario~\cite{bt1,bt2,bt3,bt4,bt5,bt6,bt7,bt8,bt9}.   These 
theories include both a  Higgs doublet $\phi$ and a technicolor sector.   Typically, the $\phi$ 
squared mass is assumed positive at the weak scale; the $\phi$ field develops a vacuum expectation 
value (vev) due to a linear term in the Higgs potential that is induced when the technifermions 
condense.   In this sense, technicolor is the trigger of electroweak symmetry breaking.  Yukawa 
couplings between $\phi$ and the quarks and leptons lead to fermion masses in the usual way.  
Since the scalar couplings to standard model fermions are the same as in a two-Higgs-doublet 
model of type I, flavor-changing neutral currents are not unacceptably large.  Moreover, it has 
been shown that ultraviolet completions exist in which bosonic technicolor with a composite Higgs 
doublet emerges as the low-energy effective theory~\cite{setc,luty}.  We will remain agnostic in 
the present work as to whether $\phi$ is fundamental or composite.

Holographic constructions of bosonic technicolor models have shown that the constraints 
on the electroweak $S$ parameter can be satisfied~\cite{cet1,carprimbtc}.  (Other discussions of 
the holographic calculation of the $S$ parameter can be found in Ref.~\cite{sss}.) In these models, 
the scales of chiral symmetry breaking and confinement can be adjusted independently.  If the 
technicolor confinement scale is chosen such that the technirho mass is kept above $\sim 1.5$~TeV, 
then one finds that the $S$ parameter constraints are satisfied over ranges of the technipion 
decay constant, $f$, that never exceed $f \sim 0.4\, v$,  where $v=246$~GeV is the electroweak 
scale (see, for example, Fig.~3 in Ref.~\cite{carprimbtc}).  Hence, with the confinement scale 
fixed, the problematic contributions to $S$ from the technicolor sector are kept under control by 
limiting the amount of electroweak symmetry breaking that originates from the technicolor condensate.

In this paper, we point out a generic consequence of a $125$~GeV Higgs boson in 
bosonic technicolor models:  the quartic coupling in the Higgs potential can run to a negative 
value at scales that are not far above the TeV scale.   As we will show, the reason for this behavior 
is that the value of the quartic coupling at the weak scale can be significantly smaller in bosonic 
technicolor models than in the standard model, assuming in both cases a $125$~GeV Higgs 
boson.  Moreover, the top quark Yukawa coupling, which drives the quartic coupling to smaller 
values in its renormalization group evolution, is larger in bosonic technicolor than in the 
standard model.  A negative quartic coupling indicates that the potential is turning over and 
will fall rapidly to values that are beneath the desired minimum.  If this happens before the cut 
off $\Lambda$ of the effective theory, then the original vacuum state will no longer be stable.  We 
will show that only a narrow region of the model parameter space is consistent
with the requirement that the quartic coupling remain positive up to a cut off $\Lambda=10$~TeV;
this region becomes even smaller for larger values of the cut off.   We also show that this parameter
region is not substantially enlarged if one allows the vacuum to be metastable 
with a lifetime that is larger than the age of the universe.   We consider the implications of these results
in light of the other important phenomenological bounds on the parameter space of the model.

Our paper is organized as follows.  In the next section we summarize the relevant effective theory.  
In  Sec.~\ref{sec:constraints}, we discuss our procedure for determining the regions of model parameter 
space that are consistent with the vacuum stability criteria, as well as the bounds from $B^0$-$\overline{B^0}$ 
mixing, light charged Higgs searches, and the requirement that electroweak symmetry breaking occurs only when a
nonvanishing technicolor condensate is present.   In Sec.~\ref{sec:results}, we discuss our results and 
the range of validity of our approximations.  In the final section, we summarize our conclusions.

\section{The Model} \label{sec:model}

The technicolor sector of the model consists of two flavors, $p$ and $m$, that transform in the
$N$-dimensional representation of the technicolor gauge group $G_{TC}$.  We assume $G_{TC}$
is asymptotically free and confining.  Under the standard model gauge symmetry, 
$\textrm{SU}(3)_C \times \textrm{SU}(2)_W \times \textrm{U}(1)_Y$,  the left-handed technifermions
transform as an SU(2)$_W$ doublet and the right-handed components as singlets,
\begin{equation}
\Upsilon_L \equiv \left( \begin{array}{c} p \\ m \end{array} \right)_L, \,\,\,\,\, p_R,\,\,\,\,\, m_R \, .
\end{equation}
Given the hypercharge assignments $Y(\Upsilon_L)=0$, $Y(p_R)=1/2$, and $Y(m_R)=-1/2$, the 
technicolor sector is free of gauge anomalies.  We assume that $N$ is even to avoid an SU(2) Witten 
anomaly.

The technifermions form a condensate that spontaneously breaks the global $\textrm{SU}(2)_L \times \textrm{SU}(2)_R$
symmetry of the technicolor sector:
\begin{equation} \label{eq:condensate}
\left< \bar pp + \bar mm \right> \neq 0 \,\,\,.
\end{equation}
A subgroup of the global chiral symmetry is gauged, corresponding to the $\textrm{SU}(2)_W \times \textrm{U}(1)_Y$ gauge symmetry of the standard model: $\textrm{SU}(2)_W$ is identified with $\textrm{SU}(2)_L$, while $\textrm{U}(1)_Y$ is identified with the third  generator of $\textrm{SU}(2)_R$.  The condensate in Eq.~(\ref{eq:condensate}) breaks $\textrm{SU}(2)_W \times \textrm{U}(1)_Y$ to $\textrm{U}(1)_{\rm EM}$, generating masses for the $W$ and $Z$ bosons.   In extended technicolor models~\cite{etc}, one would assume at this point that additional gauge interactions, spontaneously broken at a higher scale, provide dimension-six operators that couple the condensate in Eq.~(\ref{eq:condensate}) to the standard model fermions.  These operators generate ordinary fermion masses, but quite generally produce large flavor-changing neutral current effects as well.  In contrast, bosonic technicolor models include a scalar field $\phi$ that has the quantum numbers of the standard model Higgs field, {\em i.e.}, an $\textrm{SU}(2)_W$ doublet with hypercharge $Y(\phi) = 1/2$.   This choice allows Yukawa couplings of $\phi$ to
the technifermions,
\begin{equation} \label{eq:techniyuk}
\mathcal L_{\phi T} =  -\bar\Upsilon_L \tilde\phi \, h_+ p_R -  \bar\Upsilon_L\phi \, h_- m_R + {\rm h.c.},
\end{equation}
and the ordinary fermions,
\begin{equation}\label{eq:smfyuk}
\mathcal L_{\phi f} = -\bar L_L \phi h_l E_R - \bar Q_L \tilde\phi h_U U_R - \bar Q_L \phi h_D D_R + {\rm h.c.},
\end{equation}
where $\tilde\phi = i \sigma^2 \phi^*$.   While the squared mass of $\phi$, which we will call $M^2$, can have 
any sign, bosonic technicolor models typically assume $M^2>0$; in this case, electroweak symmetry breaking does 
not occur in the absence of the technicolor condensate.  By Eq.~(\ref{eq:techniyuk}),  the condensate produces a term 
linear in $\phi$ in the scalar potential, so that $\phi$ develops a vacuum expectation value.  Masses for the standard model fermions are then generated via the Yukawa couplings in Eq.~(\ref{eq:smfyuk}).

We study this model using an electroweak chiral Lagrangian, which employs a nonlinear representation of the 
Goldstone boson fields.  We let
\begin{equation}\label{eq:sigdef}
\Sigma = \exp(2 i \Pi/f), \,\,\,\,\,  \Pi = \left( \begin{array}{cc} \pi^0/2 & \pi^+/\sqrt{2} \\ \pi^-/\sqrt{2} & -\pi^0/2
\end{array}\right) \, ,
\end{equation}
where $\Pi$ represents an isotriplet of technipions, and $f$ is the technipion decay constant.  Under the 
$\textrm{SU}(2)_L \times \textrm{SU}(2)_R$ chiral symmetry, the $\Sigma$ field transforms as
\begin{equation}
\Sigma \rightarrow L\,\Sigma\, R^\dagger \,.
\end{equation}
We may consistently include the scalar doublet $\phi$ in the effective theory using the matrix representation
\begin{equation}
\Phi = \left( \begin{array}{cc} \overline{\phi^0} & \phi^+ \\ -\phi^- & \phi^0 \end{array} \right)\,,
\end{equation}
where the columns correspond to the components of the doublets
$\tilde{\phi} = (\overline{\phi^0},\,-\phi^-)^T$ and $\phi = (\phi^+,\, \phi^0)^T$, respectively, 
with superscripts indicating the electric charges. The technifermion Yukawa 
couplings can be written as
\begin{equation}
\overline{\Upsilon}_L \left( \begin{array}{cc} \overline{\phi^0} & \phi^+ \\ -\phi^- & \phi^0 \end{array} \right)
\left( \begin{array}{cc} h_+ & 0 \\ 0 & h_- \end{array} \right) \Upsilon_R \equiv \overline{\Upsilon}_L
\Phi H \Upsilon_R \, , \label{eq:newyuk}
\end{equation}
where $\Upsilon_R$ is the column vector $(p_R , m_R)^T$.   It the product $\Phi H$ transformed as
\begin{equation}
(\Phi H) \rightarrow L \, (\Phi H)\, R^\dagger \, ,
\end{equation}
then Eq.~(\ref{eq:newyuk}) would be $\textrm{SU}(2)_L \times \textrm{SU}(2)_R$ invariant.  This implies that one 
may correctly include $\Phi H$ in the effective chiral Lagrangian as a spurion with this transformation rule. The 
lowest-order term involving $\Phi H$ is
\begin{equation} \label{eq:PhiHmixing}
\mathcal L_H = c_1 4\pi f^3 \textrm{Tr}(\Phi H \Sigma^\dagger) + h.c. \; .
\end{equation}
Here $c_1$ is an unknown, dimensionless coefficient.  One would expect $c_1$ to be no smaller than 
${\cal O}(1)$ by naive dimensional analysis~\cite{Manohar:1983md}.  As in Refs.~\cite{cet1,carprimbtc}, 
we simplify the parameter space by assuming that $h_+=h_-\equiv h$, so that there is no explicit violation of 
custodial isospin from a technifermion mass splitting.

We choose to decompose $\Phi$ into its isosinglet and isotriplet components, $\sigma$ and $\Pi'$ respectively, using 
a nonlinear field redefinition similar to Eq.~(\ref{eq:sigdef}):   
\begin{equation} \label{eq:pisigma}
\Phi = \frac{\sigma+f'}{\sqrt{2}}\Sigma', \,\,\,\,\,\Sigma' = \exp(2 i \Pi'/f')\,.
\end{equation}
Here $f'$ represents the $\phi$ vev.    The kinetic terms for the scalar fields may then be expressed
\begin{equation} \label{eq:chirallagrangian}
\mathcal L_{KE} = \frac{1}{2}\partial_\mu\sigma\partial^\mu\sigma
+\frac{f^2}{4}\textrm{Tr}(D_\mu\Sigma^\dagger D^\mu\Sigma)
+\frac{(\sigma+f')^2}{4}\textrm{Tr}(D_\mu\Sigma'^\dagger D^\mu\Sigma'),
\end{equation}
where the covariant derivative is 
\begin{equation}
D^\mu\Sigma = \partial^\mu\Sigma-igW_a^\mu\frac{\tau^a}{2}\Sigma+ig'B^\mu\Sigma\frac{\tau^3}{2}.
\end{equation}
Terms in Eq.~(\ref{eq:chirallagrangian}) that mix the gauge fields with derivatives of
scalar fields allow us to identify the unphysical linear combination
\begin{equation} \label{eq:absorbedmixing}
\pi_a = \frac{f\,\Pi+f'\,\Pi'}{\sqrt{f^2+f'^2}} \, ,
\end{equation}
which is eliminated in unitary gauge. The orthogonal linear combination
\begin{equation} \label{eq:physicalmixing}
\pi_p = \frac{-f'\,\Pi+f\,\Pi'}{\sqrt{f^2+f'^2}}\, ,
\end{equation}
is physical and remains in the low-energy theory.  The mass of this multiplet follows from 
Eq.~(\ref{eq:PhiHmixing}):
\begin{equation} \label{eq:mpi}
m_\pi^2 = 8 \sqrt{2} \pi c_1 h \frac{f}{f'} v^2  \,.
\end{equation}
The masses of the $W$ and $Z$ bosons follow from Eq.~(\ref{eq:chirallagrangian})
\begin{equation}
m_W^2 = \frac{1}{4}g^2v^2,\,\,\,\,\,\,\,\,\,\,  m_Z^2=\frac{1}{4}(g^2+g'^2)v^2,
\end{equation}
where $v=246$~GeV is the electroweak scale and
\begin{equation} \label{eq:vff}
v^2 \equiv f^2+f'^2 \,\, .
\end{equation}

The coupling of the $\Phi$ field to quarks is given by  
\begin{equation} \label{eq:sigmaquark}
\mathcal L_{\Phi \bar q q}= -
\bar\psi_L \Phi \left(\begin{array}{cc} h_U&0\\ 0& V_{CKM}h_D \end{array}\right) \psi_R + \textrm{h.c.} \ ,
\end{equation}
where $\psi_L = (U_L, V_{CKM}D_L)$, $\psi_R = (U_R, D_R)$, $h_U = \textrm{diag}(h_u,h_c,h_t)$,
and $h_D = \textrm{diag}(h_d,h_s,h_b)$, or using Eq.~(\ref{eq:pisigma}),
\begin{equation}
\mathcal L_{\Phi \bar q q} = - \frac{\sigma + f'}{\sqrt{2}} \bar\psi_L \Sigma'
\left(\begin{array}{cc} h_U&0\\ 0& V_{CKM} h_D \end{array}\right) \psi_R + \textrm{h.c.}
\label{eq:pictq}
\end{equation}
The dependence of this expression on $f'$ rather than $v$ indicates that the Yukawa couplings shown 
are numerically larger than in the standard model.   In addition, Eq.~(\ref{eq:pictq}) allows one to extract that
$\bar{q} \,\Pi' q$ vertex, from which one can deduce the coupling of the physical pions $\pi_p$ to quarks.
This will be used in our subsequent phenomenological analysis.

\section{Constraints} \label{sec:constraints}

In this section, we describe our approach to studying the parameter space of the model.   We first note
that specifying $f'/v$ determines the technipion decay constant via Eq.~(\ref{eq:vff}) and, hence, the mixing angles 
that appear in Eqs.~(\ref{eq:absorbedmixing}) and (\ref{eq:physicalmixing}).  The bounds following from the virtual 
exchange or the real production of charged technipions (relevant later in this section) are then completely determined when 
$m_\pi$ is specified.  Moreover, if the technipion Yukawa coupling $h$ is not too large, then the unknown parameters
$c_1$ and $h$ appear at leading order in our vacuum stability analysis only via their product, which can be replaced 
by $m_\pi$ using Eq.~(\ref{eq:mpi}).  We therefore find it convenient to describe the model in terms of a two-dimensional
parameter space, the $f'/v$-$m_\pi$ plane.  After discussing the relevant phenomenology below, our results are presented in
Sec.~\ref{sec:results}.

\subsection{Vacuum Stability}
The form of the  scalar potential in bosonic technicolor models suggests that the requirement of vacuum stability may yield a
meaningful constraint.  (For a general review of vacuum stability bounds, see Ref.~\cite{sher}.)  Consider the potential
\begin{equation}
V(\sigma) = \frac{1}{2} M^2 \, \sigma^2 + \frac{1}{8} \lambda \, \sigma^4 \, - \frac{f^2 f'}{v^2} m_\pi^2 \, \sigma 
- \frac{3}{64 \pi^2} h_t^4 \sigma^4 \left[\ln \left(\frac{h_t^2 \sigma^2}{2\, m_Z^2}\right)-\frac{3}{2}\right] \, ,
\label{eq:thepot}
\end{equation}
renormalized at the scale $m_Z$ in the $\overline{{\rm MS}}$ scheme.  The first two terms represent the tree-level potential 
of the standard model.  The third term originates from the coupling of the Higgs boson to the technifermion condensate 
in Eq.~(\ref{eq:PhiHmixing}) and has been expressed in terms of the technipion mass.  The final term is the largest radiative 
correction, from a top quark loop.   We have checked that the radiative corrections that we omit from Eq.~(\ref{eq:thepot}) have a 
negligible effect on our numerical results, provided that $h$ is not too large.  We generally assume that
$h^2 \ll h_t^2$; we discuss this approximation further in Sec.~\ref{sec:results}.

The conditions $V_0'(f')=0$ and $V_0''(f')=m_\sigma^2$, where $m_\sigma$ is the running Higgs boson mass,  allow us to solve for the Higgs quartic coupling $\lambda$ and the Lagrangian Higgs squared mass $M^2$:
\begin{equation}
M^2 = -\frac{1}{2}\, m_\sigma^2-\frac{3}{16 \pi^2} h_t^4 \,{f'}^2+ \frac{3}{2} \, \frac{f^2}{v^2}\, m_\pi^2 \, .
\label{eq:msqsolve}
\end{equation} 
\begin{equation}
\lambda = \frac{m_\sigma^2}{{f'}^2}+\frac{3}{8\pi^2}\, h_t^4\, \ln\left(\frac{h_t^2 {f'}^2}{2 \,m_Z^2}\right)-\frac{f^2}{{f'}^2}\, \frac{m^2_{\pi}}{v^2} \, ,
\label{eq:lamsolve}
\end{equation}
Notice that the effect of the linear term in Eq.~(\ref{eq:thepot}) is to reduce $\lambda$ in Eq.~(\ref{eq:lamsolve}) relative to its 
value in the standard model.  In fact, this reduction is most pronounced when one requires $M^2>0$, since 
Eq.~(\ref{eq:msqsolve}) then implies that $f^2 m_\pi^2 / v^2$ must be non-negligible.  In any case, the running of $\lambda$ 
to higher scales is affected most strongly by the top quark Yukawa coupling,
\begin{equation}
h_t = \sqrt{2}\,\frac{m_t}{f'} \, ,
\label{eq:htmt}
\end{equation}
which is larger than in the standard model, since $f' <v$; the top quark Yukawa coupling drives $\lambda(\mu)$ to smaller values in its 
renormalization group running, where $\mu$ is the renormalization scale.   Since $\lambda(\mu)$ is smaller at $\mu=m_Z$ and the 
running of $\lambda$ is faster, one generically expects stronger vacuum stability constraints in bosonic technicolor than in the standard model. 

We consider two possible criteria for establishing the vacuum stability of the model.  We first consider the requirement that the 
quartic coupling $\lambda$ remain non-negative below a specified cut off for the low-energy effective theory, i.e., 
\begin{equation}
\lambda(\mu) \geq 0 \,\,\, \mbox{  for  } \,\,\, \mu \leq \Lambda .
\label{eq:killone}
\end{equation}  
Just beyond the scale at which $\lambda$ becomes negative, one expects the potential to turn over and drop to values 
below the minimum at $v \approx 246$~GeV.   If this occurs for $\mu > \Lambda$, one can assume that new
physics becomes relevant above the cutoff scale and alters the theory so that a deeper minimum in the potential is not obtained.  
In our numerical  analysis, we first consider the implications of this assumption for  $\Lambda=10$, $100$ and $1000$~TeV.  Since the 
LHC center-of-mass energy will not exceed $14$~TeV, and the the energies available for parton-level processes are only a fraction 
of this, our smallest choice for $\Lambda$ is still sufficient to assure that the effective theory defined in Sec.~\ref{sec:model} is the 
appropriate description of the physics that is relevant at LHC energies. 

Alternatively, one might require that the maximum of the potential occur before the cut off of the effective theory, since the 
potential drops precipitously afterwards.   Above the technicolor confinement scale, we assume the potential is given by 
Eq.~(\ref{eq:thepot}) without the linear term (since the technifermions have not yet condensed).  As discussed in the context 
of the standard model in Ref.~\cite{casas}, the maximum is reached when the 
quantity $\tilde{\lambda} \sim 0$, where  
\begin{eqnarray}
\tilde{\lambda} &=& \lambda - \frac{1}{16 \pi^2} \left\{ 6 \, h_t^4 \left[\ln\frac{h_t^2}{2}-1\right]-\frac{3}{4} \, g^4 \left[\ln\frac{g^2}{4}-\frac{1}{3}\right] \right. \nonumber \\
&-& \left. \frac{3}{8}\,  (g^2+{g'}^2)^2 \left[ \ln\frac{(g^2+{g'}^2)}{4} - \frac{1}{3}\right] \right\} \, ,
\end{eqnarray}
where $g$ and $g'$ and the standard model SU(2)$_W$ and U(1)$_Y$ gauge couplings.  We determine the model parameter space 
in which the vacuum is stable following from the criterion
\begin{equation}
\tilde{\lambda}(\mu) \geq 0 \,\,\, \mbox{  for  }\,\,\, \mu \leq \Lambda ,
\label{eq:killtwo}
\end{equation}
and compare to the results that follow from Eq.~(\ref{eq:killone}).

Finally, we consider the possibility that the potential does fall to a value lower than the desired minimum, but that the lifetime 
of the false vacuum decay is longer than the age of the universe.  In this case, the lowest point in the potential occurs at 
$\phi = \Lambda$, where new physics at the cut off may produce a second local minimum.  The requirement that the
quantum tunneling rate at zero temperature is sufficiently small may be approximated~\cite{arnold}
\begin{equation}
e^{409} \max_{\lambda(\phi)<0}\left[\left(\frac{\phi}{v}\right)^4 \exp\left(-\frac{16 \pi^2}{3 |\lambda(\phi)|}\right)\right] \alt 1  \,\, ,
\label{eq:tunnel}
\end{equation}
where we have rewritten the condition given in Ref.~\cite{arnold} in terms of our definition of the quartic coupling.  The 
quantity in brackets is maximized when $\phi=\Lambda$, where $\lambda(\phi)$ is most negative.  We will see that the model
parameter space consistent with Eq.~(\ref{eq:tunnel}) is slightly larger than what one obtains assuming Eq.~(\ref{eq:killtwo}).
Note that true vacuum bubbles may also nucleate due to thermal excitation, which typically leads to constraints intermediate
between Eqs.~(\ref{eq:killtwo}) and (\ref{eq:tunnel}); since the difference is not large in the present model, we will not consider
this issue further here.
 
Let us now summarize the fixed input parameters that are used in our analysis.  In solving for $M^2$ and $\lambda$, 
Eqs.~(\ref{eq:msqsolve}) and (\ref{eq:lamsolve}), we require the Higgs boson running mass $m_\sigma(\mu)$ and the top 
quark Yukawa coupling $h_t(\mu)$, both evaluated at  the scale $m_Z$.  The relationship between the physical Higgs boson 
mass $m_0$ and the running mass is given by~\cite{casas}
\begin{equation}
m_0^2 = m_\sigma^2(m_Z) + {\rm Re}[ \Pi(p^2 = m_0^2) - \Pi(p^2=0)] \, ,
\end{equation}
where $\Pi(p^2)$ is the renormalized self-energy of the Higgs boson; in our analysis, we include only the largest effects 
proportional to $h_t^2$, consistent with our previous approximations.  Explicit expressions for these self-energies can be
found in Ref.~\cite{ceqr}. We take $m_0=125$~GeV in determining  $m_\sigma^2(m_Z)$.   The running top quark 
mass at $m_t$ is related to the top quark pole mass ${m_t}_0=172$~GeV by
\begin{equation}
{m_t}_0 = \left[1 + \frac{4}{3} \frac{\alpha_3({m_t}_0)}{\pi}\right]\, m_t({m_t}_0) \, ,
\end{equation}
where we have taken into account the largest, QCD corrections~\cite{casas}.  With $m_t({m_t}_0)$ determined from this expression,
one uses Eq.~(\ref{eq:htmt}) to determine the running top quark Yukawa coupling evaluated at the same scale, 
$h_t({m_t}_0)$.  We then use the renormalization group equations (RGEs) to determine $h_t(m_Z)$, so that we may 
evaluate Eqs.~(\ref{eq:msqsolve}) and (\ref{eq:lamsolve}) at the same scale.

With $\lambda(m_Z)$ thus determined, we may solve the coupled one-loop RGEs for $\lambda$, $h_t$ and the standard 
model gauge couplings to determine whether the criteria in Eqs~(\ref{eq:killone}), (\ref{eq:killtwo}) and (\ref{eq:tunnel}) are met.  
We use the standard model RGEs given in the appendix of Ref.~\cite{arason}.   We have estimated the
effect of the technicolor sector on the RGE running by comparing our results to those obtained when including the 
perturbatively calculated one-technifermion-loop contribution to the standard model gauge coupling beta functions.  (All 
effects proportional to the techifermion Yukawa coupling $h$ are suppressed given our assumption that $h^2 \ll h_t^2$.)
We find that this exercise produces no noticeable effect on our results.

\subsection{$B^0$-$\overline{B^0}$  Mixing}
It is well known that $B^0$-$\overline{B^0}$ mixing provides a useful constraint on two-Higgs doublet models~\cite{hhg}.
Box diagram contributions from charged technipion exchange have also been studied in the context of bosonic technicolor models in the past (for example, in Refs.~\cite{bt2,bt3,bt4,bt7}).  Using results available in the literature on two-Higgs-doublet models, we evaluate the charged technipion contribution to $B^0$-$\overline{B^0}$ mixing, taking into account next-to-leading-order (NLO) QCD corrections.  We will see in the next section that the importance of this analysis is that the combined constraints from vacuum stability and $B^0$-$\overline{B^0}$ mixing eliminate substantial regions of the model's parameter space in which $f'$ is not close to $v$.

Extracting the charged technipion couplings to quarks from Eq.~(\ref{eq:pictq}), one finds
\begin{equation}
{\cal L} = i \, \frac{g}{\sqrt{2} m_W} \frac{f}{f'} \pi^+_p \sum_{ij} \left[
\bar{u}^i_R \, m^i_u V_{ij} \, d_L^j - \bar{u}^i_L V_{ij} m_d^j \, d_R^j \right] + \mbox{ H.c.}\, ,
\label{eq:chpicoup}
\end{equation}
where $V_{ij}$ is the Cabibbo-Kobayashi-Maskawa (CKM) matrix and the fields are given in the mass eigenstate basis.  
Since we retain only effects proportional to powers of the top quark Yukawa coupling, the term proportional to $m_d$ can 
be ignored.  Then the $\pi^+$ coupling can be matched to the charged Higgs coupling in a two-Higgs-doublet model of 
either Type I or II with the identification
\begin{equation}
\tan\beta \equiv \frac{f'}{f}  \, ,
\end{equation}
where $\tan\beta$ generally represents the ratio of the vev of the Higgs field that couples to the top quark to the vev of the Higgs field that doesn't.  In comparing the $\pi^\pm$ vertex in Eq.~(\ref{eq:chpicoup}) to the corresponding charged Higgs coupling in a two-Higgs-doublet model, an overall phase difference is irrelevant here since the diagrams of interest always connect 
each $\pi^+$ vertex to a $\pi^-$ vertex with a technipion propagator. At leading order (LO), one finds that the neutral $B$ meson mass
splitting is given by
\begin{equation}
\Delta m_B^{LO} = \frac{G_F}{6 \pi^2} m_W^2 |V_{td} V_{tb}^*|^2 f_B^2 \hat{B}_{B_d} m_B \left(
I_{WW} + \cot^2\beta \, I_{W\pi} + \cot^4\beta \, I_{\pi\pi} \right) \, ,
\label{eq:dmblo}
\end{equation}
where $f_B$ is the $B$ meson decay constant, $\hat{B}_{B_d}$ is the bag factor, and the $I_{ab}$ are given by~\cite{bbref}
\begin{eqnarray}
I_{WW}&=&\frac{x}{4} \left(1+\frac{9}{\left(1-x\right)}-\frac{6}{\left(1-x\right)^2} 
- \frac{6}{x} \left(\frac{x}{1-x}\right)^3\ln x\right) \, ,\nonumber \\
I_{W\pi}&=&\frac{xy}{4} \left[ -\frac{8-2x}{(1-x)(1-y)}+\frac{6z\ln x}{(1-x)^2(1-z)} 
+\frac{\left(2z-8\right)\ln y}{(1-y)^2(1-z)} \right] \, ,\nonumber \\
I_{\pi\pi}&=&\frac{xy}{4} \left[ \frac{\left(1+y\right)}{\left(1-y\right)^2} + \frac{2y\ln y}{\left(1-y\right)^3} \right] \, ,
\end{eqnarray}
where $x = {m_t^2}/{m_W^2}$, $y={m_t^2}/{m_{\pi}^2}$ and $z={x}/{y}={m_{\pi}^2}/{m_W^2}$. The NLO form for $\Delta m_B$ takes into
account running from the scale at which the effective $\Delta B =2$ four-fermion operators are generated, conventionally 
taken to be $m_W$, down to the $B$ meson mass scale.  The NLO expression for $\Delta m_B$ is lengthy and 
can be found in Ref.~\cite{bbref}.  We evaluate the NLO prediction assuming the lattice QCD estimate 
$f_B \sqrt{\hat{B}_d} = 216 \pm 15$~MeV~\cite{lattice}, which represents the largest source of theoretical uncertainty.

The standard approach to obtaining charged Higgs bounds from $\Delta m_B$ is to fix the CKM elements to the values
obtained in a standard model global fit.  Since such fits are consistent with the experimental data, one then requires
that the NLO prediction from $\Delta m_B$ not deviate by more than two standard deviations from the experimental value.
More precisely, we define
\begin{equation}
\chi^2 = \frac{(\Delta m_B - \Delta m_B^{{\rm exp}})^2}{\sigma^2} \, ,
\label{eq:chisq}
\end{equation}
and require that the $\chi^2$ not exceed $3.84$ to determine the 95\% confidence level (C.L.) bound.  The error $\sigma$
includes both the theory and experimental errors added in quadrature.  We take $\Delta m_B^{{\rm exp}} = 
(3.337 \pm 0.033) \times 10^{-10}$~MeV~\cite{pdg}.

\subsection{Charged Higgs Searches}

Charged Higgs searches at colliders can potentially exclude some regions of the $f'/v$-$m_\pi$ plane.  Most of the existing analyses make specific (and often simplified) assumptions about the charged Higgs decay modes and branching fractions that do not apply to bosonic technicolor models.  As the LHC extends its reach, a dedicated analysis is required to reliably determine the bounds on charged technipions in the present model.  However, for technipion masses below $m_W$ the situation is much simpler: only decays to standard model quarks (excluding the top quark) and leptons are kinematically available.   Given that the charged technipion couplings are proportional to fermion masses, as in Eq.~(\ref{eq:chpicoup}), the dominant decay channels are $\pi^+ \rightarrow \tau^+ \nu$ and $\pi^+ \rightarrow c \, \bar{s}$.  The LEP working group for Higgs boson searches has established a bound on charged Higgs bosons predicted in two-doublet extensions of the standard model, produced via $e^+e^-\rightarrow H^+ H^-$~\cite{lepwg}. The coupling of the technipions to the photon and $Z$ boson follow from Eq.~(\ref{eq:chirallagrangian}),
\begin{equation}
{\cal L}= -i\left[ e A^\mu+\frac{e}{2 s_w c_w}(c_w^2-s_w^2) Z^\mu\right] (\pi_p^+ \partial_\mu \pi_p^- - \pi_p^- \partial_\mu \pi_p^+) \, ,
\label{eq:prod}
\end{equation}
where $s_w$ ($c_w$) represents the sine (cosine) of the weak mixing angle.  Eq.~(\ref{eq:prod}) is the same as in a generic two-Higgs-doublet model (with the convention that $e$ is a negative quantity).  Hence, the production cross section for physical technipions in bosonic technicolor is consistent with the assumptions of the LEP analysis.  Moreover, this analysis assumes $\tau \nu$ and $c \bar{s}$ decays only, with arbitrary branching fractions, consistent with the present model when $m_\pi < m_W$.  Hence, the LEP lower bound of $78.6$~GeV (95\% C.L.) directly applies.  We take this into account in the following section.

\section{Results} \label{sec:results}
\begin{figure}
    \centering
        \includegraphics[width=8cm,angle=0]{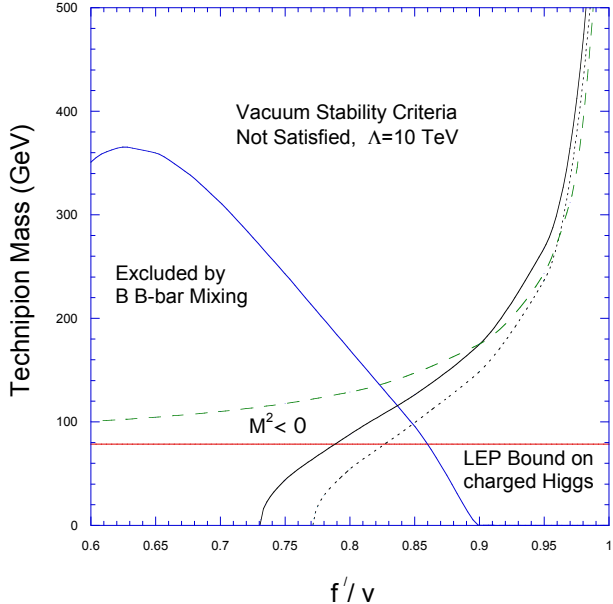}
    \caption{The model parameter space, assuming a $125$~GeV Higgs boson.  In the region above the solid [dotted] line on the 
    right, $\tilde{\lambda}(\mu) < 0$ [$\lambda(\mu) < 0$] for $\mu \le \Lambda$.  The region below the solid line on the left is 
    excluded by $B^0$-$\bar{B^0}$ mixing. The region below the horizontal solid line is excluded by the LEP charged Higgs 
    bound. The Higgs doublet squared mass is positive above the dashed line. \label{fig:xp}}
\end{figure}

The various regions of the model parameter space are displayed in Fig.~\ref{fig:xp}, for the choice of cut off
$\Lambda=10$~TeV.   Neither of the vacuum stability criteria given in Eqs.~(\ref{eq:killone}) and (\ref{eq:killtwo}) are 
satisfied  above the solid line on the right of the figure (the line that asymptotes to $f'/v \sim 0.98$).   The condition
$\lambda(\mu) \ge 0$ for $\mu \le \Lambda$ is not satisfied above the dotted line that closely tracks this boundary.
Comparing the two vacuum stability criteria,  the solid $\tilde{\lambda}(\Lambda)=0$ line gives a slightly weaker bound on 
the model parameter space.  This is consistent with the observation made in Ref.~\cite{casas}, in the context of the 
standard model, that the cutoff scale associated with vanishing $\tilde{\lambda}$ is somewhat higher than the one 
associated with  vanishing $\lambda$.    The shape of the region excluded by the vacuum stability constraint is 
also consistent with one's expectations based on Eq.~(\ref{eq:lamsolve}):  for fixed $f'$, there will be some $m_\pi$ that 
will be sufficiently large such that the last term in Eq.~(\ref{eq:lamsolve}) drives $\lambda(m_Z)$ to an unacceptably 
small initial value.  Since this last term is proportional to $f/f'$, one expects that the bound becomes weaker as $f'$ 
approaches $v$. Although the cut off of $\Lambda=10$~TeV is low, the vacuum stability constraint remains significant 
since the Eq.~(\ref{eq:lamsolve}) can lead to negative $\lambda(m_Z)$, before any RGE running, if the third term in 
Eq.~(\ref{eq:lamsolve}) is sufficiently large.

The region below the solid line toward the left side of Fig.~\ref{fig:xp} is excluded by $B^0$-$\bar{B^0}$ mixing.  For fixed 
$f'$ of intermediate size, reducing the charged technipion mass enhances the new physics contribution to $\Delta m_B$ 
until Eq.~(\ref{eq:chisq}) exceeds its 95\% C.L. value.   However, one can see from Eq.~(\ref{eq:chpicoup}) that the 
charged technipion coupling to quarks is suppressed by $f/f'$; the new physics contribution becomes irrelevant as $f'$ 
approaches $v$.   From Fig.~\ref{fig:xp}, one can see that the technipion contribution to $\Delta m_B$  becomes 
irrelevant, given the total theoretical and experimental uncertainties, when $f'$ exceeds $\sim 0.9$.   If one chooses
to impose the requirement of exact vacuum stability, then the $B^0$-$\bar{B^0}$ constraints forces  $f'/v \agt 0.84$:  only 
a relatively small fraction of electroweak symmetry breaking can originate from the technicolor condensate.    For a fixed 
technicolor confinement scale, this is the same limit in which the technicolor contribution to the electroweak $S$ parameter 
was found to be under control in Ref.~\cite{cet1}.

The LEP bound on the charged technipions, discussed in the previous section, is also displayed in Fig.~\ref{fig:xp}.  
The boundary of the stable vacuum region and the solid exclusion lines leave a roughly triangular region, above $m_\pi=78.6$~GeV 
and on the far right side of the plot.    However, within this region the Lagrangian squared mass for the Higgs doublet, $M^2$, can 
have any sign.  Of course, there is nothing physically inconsistent with electroweak symmetry breaking originating in part 
from a Higgs doublet field with a negative squared mass and in part from a fermion condensate.  We know of know argument that 
would preclude such a possibility from emerging from {\em some} ultraviolet completion.  Nevertheless, bosonic technicolor models 
have typically assumed that the Higgs doublet has a positive squared mass, so that electroweak symmetry breaking does 
not occur without the presence of the technifermion condensate. Defining the theory strictly in this way, we can exclude regions 
of parameter space in which $M^2<0$, as determined from Eq.~(\ref{eq:msqsolve}):  the excluded region lies below the dashed line in 
Fig.~\ref{fig:xp}. In this case, only a narrow strip of parameter space lies within the stable vacuum region and above the dashed 
line at which $M^2$ changes sign.  In this region, $f'/v \agt 0.9$ and the role of the technicolor condensate in electroweak symmetry 
breaking is even more limited.

Larger values of the cut off lead to more limited regions of parameter space in which Eqs.~(\ref{eq:killone}) and (\ref{eq:killtwo})
are satisfied.
\begin{figure}
    \centering
        \includegraphics[width=8cm,angle=0]{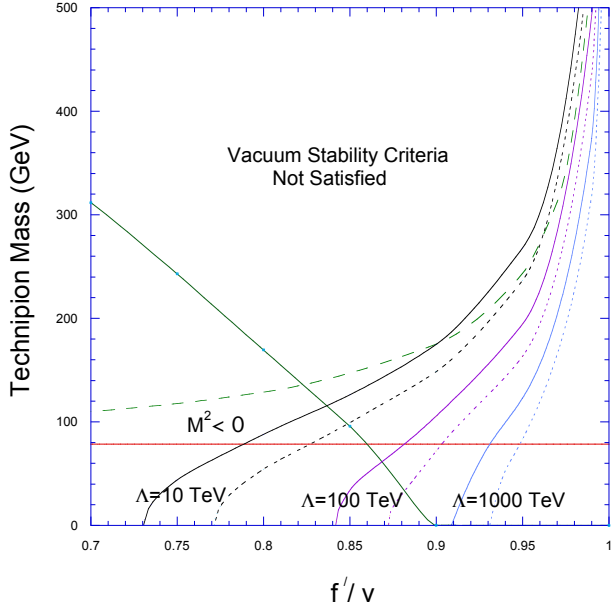}
    \caption{Vacuum stability constraints for $\Lambda=10$, $100$ and $1000$~TeV.  Otherwise,
    the lines shown have the same meaning as in Fig.~\ref{fig:xp}. \label{fig:vary}}
\end{figure} 
In Fig.~\ref{fig:vary}, we show how Fig.~\ref{fig:xp} changes as the cut off is increased from
$10$ to $100$ to $1000$~TeV.   For the higher choices of cut off, the entire region in which $M^2>0$ becomes
disjoint with the regions in which Eqs.~(\ref{eq:killone}) and  (\ref{eq:killtwo}) are satisfied.  One might argue that flavor-changing 
higher-dimension operators generated directly at the cut off scale could present phenomenological problems if this scale is much 
below $100$-$1000$~TeV.  However, without knowing what operators are actually generated when matching the effective theory 
to the ultraviolet  completion at $\Lambda$, one cannot draw a definite conclusion on the size of $\Lambda$ based on this argument.

\begin{figure}
    \centering
        \includegraphics[width=8cm,angle=0]{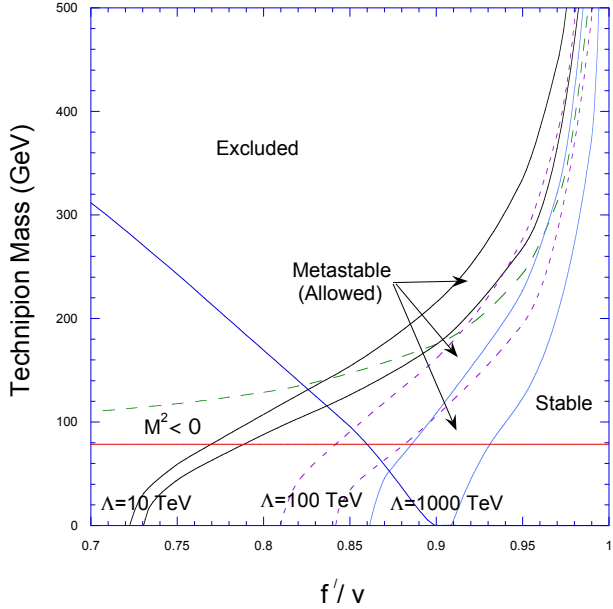}
    \caption{The model parameter space, showing bands in which the vacuum is metastable with
    a lifetime longer than the age of the universe.   From left to right, the bands with solid, dashed and solid boundary lines
    correspond to $\Lambda=10$, $100$ and $1000$~TeV, respectively.    
    \label{fig:metafig}}
\end{figure} 
In the preceding discussion, we have been careful not to describe the region in which Eq.~(\ref{eq:killtwo}) is violated
as ``excluded".   As discussed in Sec.~\ref{sec:constraints}, the model could be viable in parts of this region where the
vacuum is metastable with a lifetime that is longer than the age of the universe.  In Fig.~\ref{fig:metafig}, we show the regions
in which an acceptable metastable vacuum is obtained, following from Eq.~(\ref{eq:tunnel}), for $\Lambda=10$, $100$ and 
$1000$~TeV.    For each choice of $\Lambda$, the boundary between the given region and the one of exactly stable vacua
is given by the $\tilde{\lambda}(\Lambda)=0$ line discussed earlier.   While the excluded parameter space is somewhat smaller 
than the areas of Figs.~\ref{fig:xp} and \ref{fig:vary} in which the vacuum stability criteria are violated, these regions are not 
wildly different.  The combined constraints from $B^0$-$\bar{B^0}$ mixing and exact vacuum stability implied before that 
$f'/v \agt 0.84$; allowing for a long-lived metastable vacuum changes this inequality to  $f'/v \agt 0.825$.   Requiring that $M^2>0$ and
exact vacuum stability implied before that $f'/v \agt 0.9$; allowing for a long-lived metastable vacuum changes this to $f'/v \agt 0.835$.

Before concluding this section, we comment on the range of validity of the approximations that were assumed in this 
analysis. In our treatment of vacuum stability, we assumed $h^2 \ll h_t^2$.  In this case, we do not have to
worry about the effect of ${\cal O}(h^4)$ terms in the effective potential, or ${\cal O}(h^2 \lambda)$ terms in the RGE for
the quartic coupling.  In the regime where such terms are important, one would expect that the technifermion Yukawa
coupling, like $h_t$, should further drive the Higgs quartic coupling toward negative values.  However, a reliable 
numerical analysis is not possible (at least in the present approach) since it also depends on the running of $h$;  this is 
affected by the technicolor gauge coupling, which is nonperturbative at the TeV scale.   Hence, we do not consider this 
limit in the present analysis.   One might worry that if $h$ is bounded from above ({\em e.g.}, $h \alt 1/3$ would likely 
be sufficient for the present purposes), it might not be possible to achieve the range in technipion masses displayed
in Figs.~\ref{fig:xp} and \ref{fig:vary}.   However, the technipion mass depends on the product of the unknown
coefficient $c_1$ times $h$, as shown in Eq.~(\ref{eq:mpi}); one may increase $m_\pi$ with $h$ held fixed by 
increasing $c_1$.  This is consistent with naive dimensional analysis, which only requires that $c_1$ not be significantly 
smaller than ${\cal O}(1)$ if no fine-tuning against radiative corrections is present in the effective 
theory~\cite{Manohar:1983md}. In the holographic construction of the model,  one can compute $c_1$ directly and verify that it can be large.  This fact was illustrated in Ref.~\cite{cet1} were $\sim 1$~TeV physical technipion masses were obtained even 
with $h \sim 0.01$.   Of course, this does not imply that $c_1$ can be made arbitrarily large.   Eq.~(\ref{eq:PhiHmixing})
contains a $\pi_p^4$ vertex that is proportional to $c_1 h$.  Requiring, for example, that the ${\pi_p^+}^2{\pi_p^-}^2/4$ coupling remain
perturbative ( $< 16 \pi^2$) places an upper bound on $c_1 h$, or equivalently $m_\pi$, which we find to be
\begin{equation}
m_\pi < 2 \sqrt{6} \pi \, v \left(\frac{f'}{v}\right) \sqrt{1-\frac{{f'}^2}{v^2}} \,\,\, .
\end{equation}
For example, for $f'/v$ of $(0.9, 0.99, 0.999)$ one finds that $m_\pi$ must be less than $(1485, 528, 169)$~GeV.  Hence,
the portions of the Figs.~\ref{fig:xp}-\ref{fig:metafig} that are restricted by this perturbativity bound are at the far right edge of 
each plot and are extremely small. 

\section{Conclusions} \label{sec:conclusions}

In previous studies of bosonic technicolor models, the Higgs boson mass has been an undetermined parameter.  
Here, we have considered the consequences of fixing the Higgs boson mass at the value suggested by data from the 
2011 LHC run.   We have shown that minimization of the scalar potential in bosonic technicolor models leads to smaller 
values of the Higgs boson quartic coupling at the weak scale than in the standard model;  upon renormalization group 
running, the quartic coupling can become negative before the cut off of the low-energy effective theory, which we have 
chosen to range from $\Lambda = 10$ to $1000$~TeV.    Even with a cut off as low as $10$~TeV, we find that vacuum
stability is obtained in only a limited region of the model parameter space.   For a fixed choice of technicolor condensate, vacuum 
stability places an upper bound on the physical technipion mass, since larger technipion masses correlate with smaller values 
of the Higgs boson quartic coupling at the weak scale.  Allowing for a metastable vacuum with a lifetime longer than the
age of the universe only slightly relaxes this constraint. On the other hand, $B^0$-$\overline{B^0}$ mixing and searches for charged 
scalars at LEP place lower bounds on the technipion mass.   The parameter space that survives can be further reduced if 
one requires a positive Lagrangian squared mass of the Higgs doublet, corresponding to the scenario in which 
electroweak symmetry breaking occurs only when triggered by the existence of a technicolor condensate.  In any case,
one finds no allowed region in which the Higgs vev is less than $\sim 0.82 \, v$, where $v=246$~GeV defines 
the electroweak scale.
 
More generally, the present analysis demonstrates that electroweak symmetry breaking could include some contribution from  
strong dynamics, even if the LHC Higgs boson signal is confirmed.  However, we have shown that coupling a new strongly 
interacting sector to the Higgs potential can affect the stability of the vacuum leading to meaningful constraints on the 
allowed parameter space of such models.
 
\begin{acknowledgments}
The author thanks Marc Sher for many useful conversations on the vacuum stability bounds on Higgs potentials, and Josh Erlich for
useful comments.  This work was supported by the NSF under Grant PHY-1068008.  In addition, the author thanks Joseph J. Plumeri II
for his generous support.
\end{acknowledgments}


\end{document}